\documentclass[aps,prb,twocolumn,groupedaddress,draft,showpacs,intlimits,amsmath,amssymb,floats]{revtex4}
\usepackage{bm}
\usepackage{here}
\usepackage[final]{graphicx}
\usepackage{epsfig}
\usepackage{color}
\definecolor{DarkBlue}{rgb}{0.1,0.1,0.5}
\definecolor{Red}{rgb}{0.9,0.0,0.1}
\definecolor{Green}{rgb}{0.0,0.99,0.0}

\begin{document}
\title{Cu $K$-edge Resonant Inelastic X-Ray Scattering in Edge-Sharing Cuprates}
\date{\today}
\author{F. Vernay$^{1,2}$}
\author{B. Moritz$^{1,2}$}
\author{I. S. Elfimov$^{3}$}
\author{J. Geck$^{3}$}
\author{D. Hawthorn$^{3}$}
\author{T. P. Devereaux$^{1,2}$}
\author{G. A. Sawatzky$^{2,3}$}
\affiliation{$^{1}$
Department of Physics and Astronomy,
University of Waterloo, Waterloo, Ontario N2L
3GI, Canada} \affiliation{$^{2}$Pacific Institute of Theoretical
Physics, University of British Columbia, Vancouver BC}
\affiliation{$^{3}$Department of Physics, University of British
Columbia, Vancouver, BC}

\begin{abstract}
We present calculations for resonant inelastic x-ray scattering
(RIXS) in edge-shared copper oxide systems, such as
Li$_2$CuO$_2$ and CuGeO$_{3}$, appropriate for hard x-ray scattering
such as the copper $K$-edge. We perform exact diagonalizations of the
multi-band Hubbard model
and determine the energies, orbital
character and resonance profiles of excitations which can be
probed via RIXS. We find excellent agreement with recent results
on Li$_{2}$CuO$_{2}$ and CuGeO$_{3}$ in the 2-7 eV photon energy
loss range.

\end{abstract}
\maketitle

\section{Introduction}
Techniques that probe the momentum dependence of strongly
correlated electrons provide valuable information that can be used
as a test for models appropriate for the cuprates. Angle-resolved
photoemission (ARPES) provides important information about
single-particle excitations of occupied states.\cite{ARPES-RMP,ARPES}
Electronic Raman scattering reveals
information about multiparticle excitations at long wavelengths.
\cite{Raman-RMP} Resonant inelastic x-ray scattering (RIXS),
a form of Raman spectroscopy,
recently has undergone vast improvements, allowing probes of
momentum dependent excitations into unoccupied states.
\cite{RIXS-RMP} Improvements in resolution now offer the
possibility of a direct comparison of ARPES, Raman, and RIXS
spectra in the several eV range with the hope of understanding the
nature of strong correlations in the cuprates, and high
temperature superconductivity, by studying electronic excitations
at several energy scales.

As a fundamental building block of low-energy effective theories
of the cuprates, the Zhang-Rice singlet (ZRS) \cite{ZR} is the
starting point of the single-band Hubbard and t-J models widely
believed to capture the low energy physics of strongly correlated
systems. It is well-known that the ZRS features prominently in
RIXS in the cuprates. A particular test of models of the cuprates
lies in the difference between RIXS in edge-shared versus
corner-shared copper oxide systems. In materials such as
La$_2$CuO$_4$ and other corner sharing 2D members of the high
T$_c$ family, the ZRS is stabilized by a strong gain in
antiferromagnetic exchange energy between Cu and O, and can
propagate to neighboring CuO$_4$ plaquettes, with effective
hopping $t=-0.35$ eV along the Cu-O bond direction, and
$t^{\prime}=0.15$ eV along the diagonal.\cite{ZR} In edge-shared
cuprates, such as Li$_2$CuO$_2$ and CuGeO$_{3}$, the Cu-O-Cu
bond approximately forms a $90^{\circ}$-angle,\cite{Mizuno}
compared to $180^{\circ}$ in corner-shared cuprates, and
therefore according to Goodenough-Kanamori-Anderson,\cite{GKA}
the nearest-neighbor copper-copper exchange tends to be slightly
ferromagnetic. However, higher order processes like Cu-O-O-Cu
hopping also contribute comparable antiferromagnetic exchange
terms. For these reasons, small bond-angle variations among
different edge-sharing compounds can make the exchange slightly
anti-ferromagnetic or ferromagnetic. In either case, the ZRS is
less stable and mobile due to hybridization among different
oxygen orbitals not aligned with Cu $d_{x^2-y^2}$ orbitals.
\cite{Mizuno}

Recent Cu $K-$ edge RIXS data on edge-sharing cuprate
Li$_{2}$CuO$_{2}$ \cite{LiCuO,both} has revealed a small,
non-dispersive peak at 2.1 eV, attributed to interatomic $d-d$
excitations, a strong peak at 5.4 eV, and a weak peak at 7.6 eV,
both attributed to charge transfer excitations. The peak at 5.4
eV is doubly resonant for incident photon energies near 8986 and
9000 eV.\cite{LiCuO}  In CuGeO$_{3}$, clear excitations were
found are 3.8 eV (weak) and 6 eV (strong),\cite{both,CuGeO,Suga}
and more recently 1.7 eV,\cite{CuGeO_recent} which were
attributed in Ref. \onlinecite{CuGeO_recent} as a intra-atomic $d-d$
excitation (1.7 eV), a ZRS excitation (3.8 eV), and a charge
transfer excitation on a single copper oxide plaquette (6.4 eV).
Similar features have also appeared in oxygen 1s RIXS.
\cite{Duda} While a few momentum-dependent studies have been
performed, the incident photon energy dependence, or resonance
profile, is usually shown only for a few selected points in the
Brillouin zone.

In this paper we present a theory for RIXS for the copper
$K$-edge and present
exact diagonalizations of Hubbard
clusters of edge-sharing copper oxide plaquettes shown in Fig.~
\ref{cluster}. We calculate both the RIXS spectrum and the
resonance profile for several cluster geometries, and determine
the nature of excitations accessible via light scattering. We pay
specific attention to the intermediate core-hole and final
excited-state wavefunctions of RIXS accessible states. We find
good agreement with the results on edge-shared copper oxide
systems, and confirm a number of previous excitation assignments.
\cite{Suga,CuGeO,CuGeO_recent}

\section{Model and Method}
\subsection{RIXS process}
The RIXS process consists in scattering an incident photon of energy
$\hbar\omega_I$ off a sample in the ground state $|\psi_0\rangle$
having energy $E_0$. For $K-$edge RIXS, an electron from the $1s$
copper-core is photoexcited into the Cu ${4p}$-band, leaving
behind a core hole. The strong Coulomb repulsion between the core
hole and $d_{x^2-y^2}$ holes on the same Cu atom reorganizes
charge density, forming an intermediate state $|\psi_{ci}\rangle$
of eigenenergy $E_{ci}$. The $4p$ electron resides in a rather extended
wave-function and therefore interacts weakly with $d$ and core
states. We neglect its interaction hence providing us with a rather simple
description of RIXS. The $4p$ electrons then enter into the problem as
a spectator at energy $\epsilon_{4p}=\hbar\omega_I -(E_{ci}-E_0)$ and
the intensity will be proportional to the projected $4p$-band density
of states (DOS) at that energy. The incident polarization determines the
necessary orbital projection. The $4p$ electron then recombines with the
$1s$ core hole emitting a photon of energy $\hbar\omega_F$,
leaving the
system in the final state $|\psi_f\rangle$ of eigenenergy $E_F$ and
a photon energy transfer $\hbar\Omega=\hbar\omega_I-\hbar\omega_F$.
This leads to a simplification of the general expression given in
Ref.\onlinecite{RIXS-RMP}.
In this approximation the spectrum will be given by the following
relation and with the $\hbar\omega_I$ dependence given by a
convolution with the $4p$ band DOS~:
\begin{widetext}
\begin{equation}\label{rixs-formula}
I(\omega_I,\Omega=\omega_I-\omega_F)\propto
\sum_f\left|\sum_i\sum_{4p} \frac{\langle
\psi_f|\psi_{ci}\rangle\langle \psi_{ci}|\psi_0\rangle}
{E_{ci}+\epsilon_{4p-1s}-E_0-\hbar\omega_I-i\Gamma_1}
\right|^2\times \delta\left(E_F-E_I-\hbar\Omega\right).
\end{equation}
\end{widetext}
Here $\epsilon_{4p-1s}$ represents the Cu $1s-4p$ energy
separation. For simplicity, in a first approach,
we neglect specific $4p$ orientation and photon polarizations.
The polarization dependence, as well as the related $4p$
  orientation and the $4p$ projected DOS,
will be discussed in the last part of Section \ref{section_res}.
The parameter $\Gamma_{1}$ represents
damping of the intermediate state due to fluorescence and Auger
$1s$ decay, which we take as 1.0 eV. While this value determines
the overall resonant enhancement, it does not affect the
resolution of the energy-loss peaks in the spectrum. We broaden
the delta function with a width 0.1eV to represent current
instrument resolution. We note that in Eq.(\ref{rixs-formula}) the states
$|\psi_{ci}\rangle$ are identical to those in a core x-ray
photoemission experiment.

The Hamiltonian describing the open boundary cluster shown in Fig.~
(\ref{cluster}), with rotated $x$ and $y$ local-directions along the $p$
orbitals, is given by ${\mathcal H}=\sum_{\langle i,j\rangle
,\sigma} {\mathcal H}_{ij\sigma}^K +\sum_i (\sum_{\sigma}{\mathcal
H}^\epsilon_i+{\mathcal H}_i^U)+U_{Q}
\sum_{\sigma,\sigma^{\prime}}d_{0,\sigma}^{\dagger}d_{0,\sigma}
[1-s_{0,\sigma^{\prime}}^{\dagger}s_{0,\sigma^{\prime}}]$, with
\begin{eqnarray}\label{hamilt}
{\mathcal H}_{ij\sigma}^K &=& t_{pd}P_{i,j}\ p^\dagger_{i,\sigma}
d_{j,\sigma} + t_{pd_z}P^z_{i,j}\ p^\dagger_{i,\sigma} d_{z,\
j,\sigma}\nonumber\\
&& +t_{pp}P^{\prime}_{i,j}\ p^\dagger_{i,\sigma}p_{j,\sigma}+
t_{pp}^{\prime}\
p^{\dagger\prime}_{i,\sigma}p_{j,\sigma}+ h.c\nonumber\\
{\mathcal H}^\epsilon_i &=& \epsilon_{p}\
(p^\dagger_{i\sigma}p_{i,\sigma}+
p^{\dagger\prime}_{i,\sigma}p_{i,\sigma}^{\prime})+
\epsilon_{d}d^\dagger_{i,\sigma}d_{i,\sigma}+
\epsilon_{d_z}d^\dagger_{z_{i,\sigma}}d_{z_{i,\sigma}}\nonumber\\
{\mathcal H}_i^U &=& U_{pp}\
p^\dagger_{i,\uparrow}p_{i,\uparrow}p^\dagger_{i,\downarrow}p_{i,\downarrow}
+U_{pp}\
p^{\dagger\prime}_{i,\uparrow}p_{i,\uparrow}^{\prime}p^{\dagger\prime}_{i,\downarrow}
p_{i,\downarrow}^{\prime}\nonumber\\
&&+U_{dd}\ d^\dagger_{i,\uparrow}d_{i,\uparrow}
d^\dagger_{i,\downarrow}d_{i,\downarrow}+ U_{dd}\
d^\dagger_{z_{i,\uparrow}}d_{z_{i,\uparrow}}
d^\dagger_{z_{i,\downarrow}}d_{z_{i,\downarrow}}. \nonumber
\end{eqnarray}
Here $d^{\dagger}_{i,\sigma}, d^{\dagger}_{z_{i,\sigma}},
p^{\dagger}_{i,\sigma}, p^{\dagger\prime}_{i,\sigma}$ and
$d_{i,\sigma}, d_{z_{i,\sigma}}, p_{i,\sigma},
p_{i,\sigma}^{\prime}$ creates and annihilates a
$d_{x^{2}-y^{2}}$, $d_{3z^2-r^2}$, planar $p$ or $p'$ hole with
spin $\sigma$ at site $i$, respectively, and the sum runs over
nearest neighbors. The hopping amplitude $t_{pp}$ refers to the
hopping between two nearest $p_{x}$ $p_{y}$ orbitals,
and $t_{pp}^{\prime}$ refers to the hopping
between two nearest-neighbor $p_{\alpha}$ oxygen orbitals, where
$\alpha=x,y$. The overlap phase factors are chosen such that
$P_{i,j}=1$ for neighbors in the ${\bf x, -y}$ direction, and -1
for ${\bf -x,y}$. The oxygen phase factors are analogously
defined as $P_{i,j}^{\prime}=1$ for oxygens at ${\bf x,y}$ and
${\bf -x,-y}$, and -1 for the other combinations. Here $s_{0}^\dagger$
(resp. $s_{0}$) creates (annihilates) a Cu$_{1s}$
core-electron, and $U_Q$ is the core - $d$ hole repulsion. For
simplicity we neglect interactions involving the $4p$
photo-excited electron, such as the $4p-3d$ Coulomb interactions.
These include off-diagonal couplings between $d$ states which may
be important for singlet-triplet energy differences as well as
providing pathways for symmetry-allowed $d-d$ excitations. We
also neglect exchange interactions involving the 1s core-hole,
which are very small.

In what follows, we use a standard set of parameters [in eV]
\cite{ZR}: $t_{pd}=1.1, t_{pp}=0.5, t_{pp}^{\prime}=-0.538\times
t_{pp}, \epsilon_d=0, \epsilon_{dz}=1.7, \Delta=\epsilon_p
-\epsilon_d = 3.5, U_{pp}=6, U_{dd}=8.8$, and $U_{Q}= 8$. These
parameters give a value for the copper magnetic exchange $J=0.14$
eV for corner-shared copper oxides, and 2.8 meV for edge-shared.

\subsection{Discussion of the method and comparison with previous works}
We first perform full exact-diagonalization of the Hamiltonian without
the core-hole, and then redo the calculation with the core-hole.
Since the Hamiltonian conserves the total spin, we diagonalized
in the S=1/2 sector. From the eigenvalues and eigenvectors of the
two steps, the RIXS spectrum for a given $\hbar\omega_I$
is computed from Eq.(\ref{rixs-formula}).

As our calculations involve determining the eigenvector of each
state used to construct the matrix elements involved in RIXS, we know
the character of each excitation exactly and do not need to make a
priori assumptions about which peak corresponds to which fundamental
excitation. This allows us to make direct correspondance
between RIXS and other core-level spectroscopic techniques,
without any further assumptions.

This is quite different from the Lanczos method, which truncates the
Hilbert space and focuses on the low-lying energy excitations.
Whereas this method allows investigation of slightly larger clusters,
extensions to higher energies is problematic particularly for RIXS
where eigenvectors are needed to evaluate the matrix elements between
higher energy states in order to obtain both the resonance profile
(dependence on incident photon energies) as well as the intensities of
prominent peaks in the spectrum. Since the core-hole interaction is
essentially in nature (on-site Coulomb repulsion),
leading to only a slight delocalization of the charge,
we checked that the cluster-size does not play a crucial role in the
qualitative shape of the spectra.
In that sense, the calculations done in the present paper add more
information and provide a more direct access to the relevant
RIXS excitations compared to other techniques used in previous studies in
the context of standard cuprates.\cite{maekawa_group}
These studies treated the one-band
Hubbard model where the Zhang-Rice singlets are, by construction,
already formed, and focused only on the momentum dependence of the Mott gap
excitation. By treating the problem in a multi-band model, the
formation of Zhang-Rice excitations comes naturally in the problem and
its signature in the RIXS intensity, peak dispersion and resonance
profile can be directly addressed. In fact it is
widely recognized that the gap is a charge transfer gap which involves
transition from oxygen $2p$ to copper $3d$ states.\cite{ZSA}

\section{Results and Discussions}\label{section_res}
\subsection{Preliminary remarks}

\begin{figure}[t]
\includegraphics*[width=5.0cm]{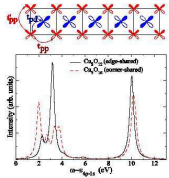}
\caption[figure1]{Top: Edge-shared Cu$_{5}$O$_{12}$ cluster.
Bottom: XPS spectrum as a function of photon energy transfer and
incident photon energy, for corner and edge-sharing copper-oxide
clusters.} \label{cluster}\end{figure}

In core level x-ray photoelectron spectroscopy (XPS), the final
state corresponds to an intermediate state in the RIXS process. We
first plot the XPS spectrum for both edge-shared cluster
Cu$_{5}$O$_{12}$ and corner-shared cluster Cu$_{5}$O$_{16}$ in
Fig.~(\ref{cluster}). In both cases, the prominent features
correspond largely to 1) a non-locally screened d-intersite ZRS
$(d^9L)$ feature at 2.2 eV, where the core hole has pushed a
$d^9$ hole from the on-site Cu onto the oxygen ligands in the
neighboring plaquette, 2) a local charge transfer (CT) excitation
$(d^{10}L)$ at 3.5 eV whereby the screening of the core hole
comes from the hole transfer from copper to oxygen
on the same plaquette,\cite{vanveenendaal} and 3) a 10 eV high-energy poorly
screened $d^9$ Cu hole on the core-hole site. These results are
quantitatively similar to those presented earlier,
\cite{early_XPS} apart from a more realistic value of the
exchange $J$ in the corner-shared geometry. We note that x-ray
absorption spectra in an extended system can be obtained from our
XPS spectrum via a convolution with the copper $4p$ DOS,
giving a broad absorption edge followed by spectral
intensity extending higher in energy governed by the $4p$
bandwidth.

These excitations, present in both corner and edge-shared
geometries, occur at roughly the same energy since they are
governed by charge transfer energies and the exchange $\propto
t_{pd}^2/\Delta$ between copper and oxygen: the Cu-Cu exchange
does not play a major role in setting these energy scales.

The main difference concerns the prominence of the non-locally
screened d-intersite ZRS excitation in the corner-shared compared
with the edge-shared systems. While the ZRS accounts for nearly 80
percent of the ground state wavefunction including the core hole
for the corner-shared plaquettes, it is approximately 56 \% for
the edge-shared system, reflecting the number of oyxgen
hybridization pathways available in the edge-shared configuration.
\cite{nb} Thus while the energy is comparable in both systems,
the matrix elements are much weaker at the ZRS scale in the
edge-shared compared to corner-shared systems. In addition, the
mobility of the ZRS for each geometry determines the peak widths,
reflecting the splitting between bonding and anti-bonding ZRS on
sites neighboring the core hole. In the case of the corner-shared
systems, this corresponds to $2t=-0.7eV$, while due to the
orientation of the orbitals in the edge-shared systems, the
splitting is $2t'=0.3$ eV. As a result, the locally screened
$d^{10}L$ peak at 3.1 eV is much larger in edge-shared compared
to corner-shared. This has led the phenomenon of high-temperature
superconductivity, occurring in corner-shared compounds and absent
in edge-shared compounds, to be associated with the stability of
the ZRS.\cite{Ohta} We note that the primary role of the oxygen
orientation in the formation of the ZRS is not captured in single-band
Hubbard model calculations.

\subsection{RIXS results~: Attribution of the peaks}
\begin{figure}[t]
\includegraphics*[width=\columnwidth,angle=0]{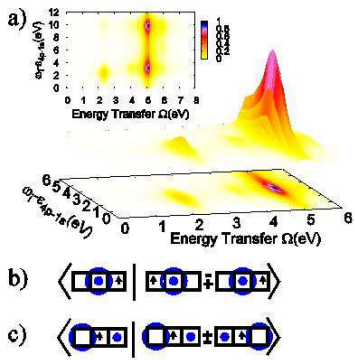}
\caption{a) RIXS spectra for Cu$_3$O$_8$ cluster as a function of
photon energy transfer $\Omega$ and incident photon energy
$\omega_I$, measured relative to the binding energy
$\epsilon_{4p-1s}$. The inset shows a contour map over large
frequencies. Matrix elements for Cu$_3$O$_8$ are shown in b) \&
c), for the main peaks in RIXS at
$(\omega_{i}-\epsilon_{4p-1s},\Omega)=(2.2,2.2)$ and $(3.1,5.07)$,
eV respectively. Here the left, right eigenvector is the core
hole, final state, respectively, and the dot and the circle
represent a singlet state formed by a hole on the copper site
(dot) and a hole delocalized on the oxygens having the ZR
symmetry (circle). } \label{wave1}
\end{figure}

The aim of this subsection is to discuss the overall shape
of the RIXS spectrum and resonant profile for a
Cu$_{3}$O$_{8}$ cluster in the inset of Fig.~\ref{wave1}a.
We show how to attribute to each peak the corresponding excitation,
 as an example, we have chosen the same parameters as for XPS in Fig.~
\ref{cluster}, appropriate for Li$_2$CuO$_2$.
From here on we set $\hbar$ equal to 1. Here the
intermediate state core-hole is included on the edge of the
cluster.
The calculations are also performed for different locations of the
core hole on the cluster, which has one hole per copper-plaquette.
While the results differ slightly for different locations of the core hole,
we note that the changes are only quantitative and give different
broadening of the RIXS peaks due to the different mobility of the
d-intersite Zhang-Rice excitation on the end or middle of the
cluster. However, since the aim of this paper is to compare trends for
different cluster geometries and charge transfer energies, we will not
discuss further these relatively minor changes in the spectra.

The RIXS spectrum shown in the inset of Fig.~\ref{wave1}a
consists of two prominent excitations at approximately
$\Omega=5.07$ eV energy transfer, at incident photon energies of
3.1 and 10 eV. This is the local CT excitation, resonant at both
the locally screened $d^{10}L$ CT and poorly screened $d^9$
intermediate states, as shown in Fig.~\ref{cluster}. In addition
a weaker d-intersite ZRS-derived peak is present at $\Omega=2.2$
eV, resonant for incident photon energies near 2.2 eV above the
$1s-4p$ transition. The XPS spectra may be crudely viewed as a
sum over all Raman shifts of the RIXS spectrum as a function
of the incident photon energy $\hbar\omega_I$.

In Fig.~\ref{wave1}a we show a blow up of the low energy RIXS
spectrum. By examining the symmetries of the core-hole and final
state wavefunctions, the nature of the RIXS excitations, and the
origin of their resonant profile, can be determined by inspection
of matrix elements appearing in Eq. \ref{rixs-formula}. The
ground state without the core hole $\mid\psi_0\rangle$ is largely
of $d^9$ character in each plaquette. Only those intermediate
states which have a large overlap with the ground state are
relevant to the RIXS process. To determine which final states
$\mid\psi_{i\ne 0}\rangle$ are probed by RIXS and at what
resonant energy they occur, one inspects the overlap of the core
hole states $\mid\psi_c\rangle$ reachable via the ground state
without the core hole $\mid\psi_0\rangle$ , with other excited
states $\mid\psi_{i \ne 0}\rangle$ in the same Hilbert space. For
example, since the ground state of the cluster
$\mid\psi_0\rangle$ is a singlet, it cannot be coupled to triplet
intermediate states having a core hole. We immediately conclude
that for our cluster all triplet states do not appear in the RIXS
process without spin-orbit coupling.

We sketch schematically in Fig.~\ref{wave1}b and \ref{wave1}c the
leading order contributions to the wave functions of the
intermediate state with the core-hole and the final state with
large matrix overlap, which correspond to the main excitations
shown in the RIXS spectra. While the intermediate state is
predominantly one configuration, the final states can either be
bonding or anti-bonding combinations of excitations on the
plaquettes, at energies approximately 2 eV above $E_{0}$, split
by ZRS hopping.

In Fig.~\ref{wave1}b, it is clear that the resonance peak at
$\omega-\epsilon_{4p-1s}=2.2eV$ and energy transfer
$\Omega=2.2eV$ comes from the combination of intermediate and
final states corresponding to the creation of the ZRS in the
final state, having strong overlap with the non-local,
well-screened $d^9L$ intermediate state. The two final states of
bonding and anti-bonding combinations of ZRS on the two
plaquettes couple to the intermediate state, with a weak
splitting in edge-shared systems. Even with our broadening of
$0.1eV$ the separate peaks are difficult to resolve.

The final states of the higher energy peak correspond to bonding
and anti-bonding combinations of $d^{10}L$, having strong overlap
with the intermediate core hole state with the ligand on the core
hole site as shown in Fig.~\ref{wave1}c. This lies at higher
energy $\omega_{i}-\epsilon_{4p-1s}=3.07eV$, and is the most
prominent final state in XPS shown in Fig.~\ref{cluster}.

We note that we do not find any prominent $d-d$ excitations lying in
the relevant energy range below 2 eV, even when including other
$d$ orbitals.\cite{dd} This follows from a symmetry analysis of
the states, whereby $d^9L$ and $d^{10}L$ configurations cannot
hybridize other Cu $d$ orbitals and combinations of oxygen
ligands. Thus direct $d-d$ excitations are inaccessible in our
model, in agreement with electron energy loss cluster calculations.
\cite{Becker} It is plausible that $d-d$ excitations may arise
if symmetry-breaking interactions with the 4p are included. This
is different for oxygen 1s RIXS, as the creation of the core hole
breaks this symmetry, allowing direct $d-d$ excitations, as
determined in recent cluster calculations.\cite{Okada}
Moreover, $d-d$ excitations may arise from multiple couplings
in non-resonant scattering\cite{Haverkort} which are not treated
in our calculation.

\subsection{Influence of the $4p$-States -- Polarization dependence
-- Material dependence}
\label{convo}
So far, our analysis did not include the copper-$4p$ states. However,
in order to obtain results more closely connected to experiment and to
include a photon polarization dependence for the Cu $K$-edge process,
the Cu-$4p$ DOS must be included into the
calculations. Indeed, the Cu-$1s$ core-electron can only be excited
into the $4p$-levels if there is a finite DOS for unoccupied states.
Due to crystal field
effects, the 4p DOS for different $4p_{x,y,z}$ orbitals provide a
polarization dependent shift of the resonance energies and secondary
satellites governed by the peaks of each $4p$ projected DOS as well as
the overall bandwidth.
The orientation of the incident and emitted photon polarizations
determine which $4p$ projected DOS is accessed.

Since the $4p$-states are rather extended orbitals their
interaction with the Hubbard subbands is quite weak. However,
depending on the incoming light-polarization, matrix elements are
selected so that one can access different linear combinations of
$4p_{x}$, $4p_{y}$ and $4p_{z}$ states. Since the DOS
distributions are differents, the RIXS spectra reflect these
characteristics. Thus to deduce the influence of the $4p$-states,
we neglect minor real-space variations of the combined
cluster-$4p$ wavefunction and compute the convolution of our
raw-RIXS spectrum with the calculated $4p$ projected DOS
$N_{4p}(\epsilon_{4p})$:

\begin{widetext}
\begin{eqnarray}\label{rixs-conv-formula}
I(\omega_I,\Omega = \omega_I-\omega_F)&\propto &
\sum_f\left|\sum_i\int d\epsilon_{4p}
N_{4p}(\epsilon_{4p})\frac{\langle \psi_f|\psi_{ci}\rangle\langle
\psi_{ci}|\psi_0\rangle}
{E_{ci}+\epsilon_{4p-1s}-E_0-\hbar\omega_I-i\Gamma_1}
\right|^2\times \delta\left(E_F-E_I-\hbar\Omega\right),\nonumber\\
&=&\sum_f\left|\int d\epsilon_{4p}
N_{4p}(\epsilon_{4p})M(\hbar\omega_I-\epsilon_{4p-1s},f)
\right|^2\times
\delta\left(E_F-E_I-\hbar\Omega\right),\nonumber\\
{\rm where}&& M(\hbar\omega_I,f)=\sum_i\frac{\langle
\psi_f|\psi_{ci}\rangle\langle \psi_{ci}|\psi_0\rangle}
{E_{ci}-E_0-\hbar\omega_I-i\Gamma_1}.
\end{eqnarray}
\end{widetext}

The projected $4p$ DOS is calculated using linearized augmented
plane wave method (LAPW) in a coordinate system consistent with
experiment.\cite{wien2k} In the following subsection we present
results for different edge-shared cuprates Li$_2$CuO$_2$ and
CuGeO$_3$.

\subsubsection{Li$_2$CuO$_2$}

\begin{figure}[t]
\includegraphics*[width=\columnwidth,angle=0]{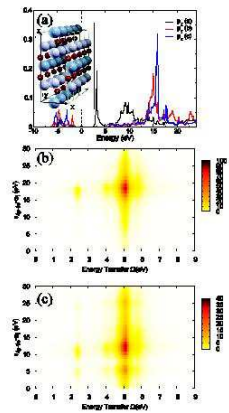}
\caption{{\bf (a)} Cu-$4p$ partial density of states for
Li$_2$CuO$_2$.
{\bf (b)} RIXS spectrum for parameters corresponding to
Li$_2$CuO$_2$, convoluted with the $p_{z}$-DOS.
{\bf (c)} RIXS spectrum for parameters corresponding to
Li$_2$CuO$_2$, convoluted with the $p_{x}$-DOS.
For both (b) and (c) we took a core-hole lifetime of
$\Gamma=1.5eV$} \label{Liconv}
\end{figure}

For Li$_2$CuO$_2$, the $4p$ projected DOS is shown in Fig.~\ref{Liconv}(a).
The $4p_{z}$ DOS has two sharp peaks at roughly
$17$eV, with subdominant peaks as shoulders to the main peak . The
resulting RIXS spectrum for $z$ photon polarizations shown in
Fig.~\ref{Liconv}(b) thus has a shifted d-intersite ZRS resonance at
$\epsilon_{4p}-\omega_i\sim 19$eV
instead of $2.2$eV in Fig.~\ref{wave1}, yet remains sharply peaked
even though the $4p_{z}$ bandwidth is over 20 eV wide. The
$d^{10}L$ CT and poorly screened excitation at $\Omega\sim 5eV$ are
likewise shifted to higher resonance frequencies.

The $4p_x$ DOS in Fig.~\ref{Liconv}(a) has a double-peak structure,
with a single sharp peak at 4 eV and a tightly structured set of peaks
around 10 eV. The resulting RIXS spectrum for photon polarizations
along the $x$-axis, given
in Fig.~\ref{Liconv}(c), clearly shows two split resonances at
$\sim 6-7$eV and $\sim 13$eV, a shift of the raw RIXS spectrum
(Fig.~\ref{wave1}) by the $4$eV and 10$eV$ $4p_x$
DOS peaks. The difference in the relative intensities between
Fig.~\ref{Liconv}(b) and Fig.~\ref{Liconv}(c) is due to the fact that the DOS
for the $p_{z}$ orbitals
is more intense at its maximum than for the $p_x$ orbital.

\begin{figure}[t]
\includegraphics*[width=\columnwidth,angle=0]{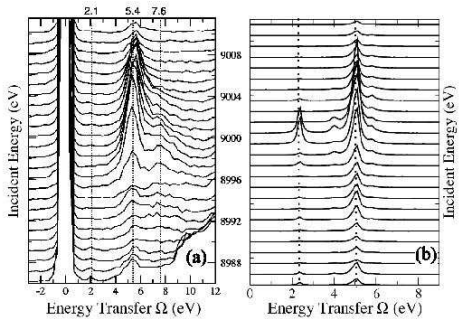}
\caption{{\bf (a)} Experimental data for Li$_2$CuO$_2$ by Kim
{\it et al.} in Ref.\onlinecite{LiCuO}.
{\bf (b)} Calculated RIXS spectrum for parameters corresponding to
Li$_2$CuO$_2$ convoluted with the $4p$-DOS.} \label{Li_comp}
\end{figure}

Fig.~\ref{Li_comp} displays a side-by-side comparison of the
present theory and recent experimental data by Kim {\it et
  al.}\cite{LiCuO} on Li$_2$CuO$_2$. The theory and
experimental curves show striking similarities in both the energy
transfer and the incoming photon energy range.
The energy scale, prominence of the local CT excitation, and the
double resonance profile shown in Figs.~\ref{cluster},\ref{wave1},\ref{Li_comp}
correspond well to the RIXS experiments. We remark that while
Ref. \onlinecite{LiCuO} attributed the
weak 2.1 eV peak to an interatomic $d-d$ excitation, and the
strong peak at 5.4 eV peak to local CT excitation, in view of our
results we would conclude that the low energy excitation rather
corresponds to the d-intersite ZRS excitation. The weak peak
observed at 7.6 eV however has no correspondance in our
calculations. This peak may be associated with a local CT
excitation into states with different ligand symmetry orthogonal
to the ZRS ligand states, which may become accessible once oxygen
$4p-2p$ interactions are included.

\subsubsection{CuGeO$_3$}

As pointed out in Ref.\onlinecite{Mizuno},
the large valency of Ge$^{4+}$ nearby the
CuO planes increases the Madelung energy difference between the
Cu and planar O sites in CuGeO$_3$ compared to LiCuO$_2$, leading
to the values $\Delta=4.9$ and 3.2 eV, respectively, for each
system. Since this energy scale largely determines the energies
of the d-intersite ZRS and the separation in energy to the local CT
excitation, the RIXS spectrum and resonance profile should be
largely different in the two cases.

\begin{figure}[t]
\includegraphics*[width=\columnwidth,angle=0]{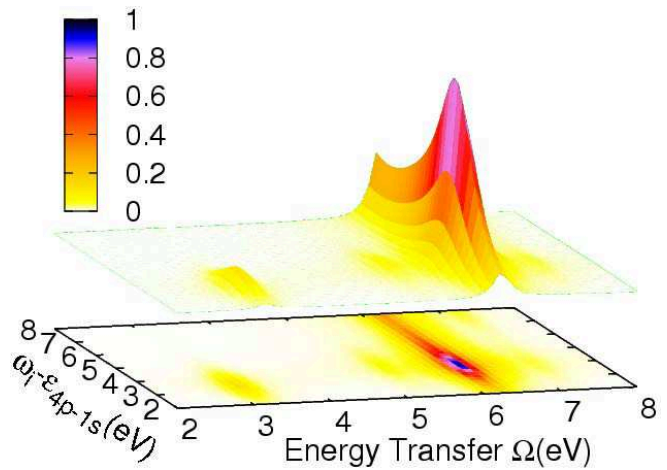}
\caption{RIXS spectrum as a function of photon energy transfer and
incident photon energy for parameters corresponding to
CuGeO$_{3}$, as defined in the text.} \label{Ge}
\end{figure}

In a first approach, as we did for Li$_2$CuO$_2$,
we present in Fig.~\ref{Ge} a non-convoluted RIXS spectrum for
parameters corresponding to CuGeO$_{3}$, which are the same set
of parameters as before apart from the change $t_{pd}$=1.23 eV,
and $\Delta=4.9$ eV. The larger value of $\Delta$ leads to
several changes compared to the RIXS spectrum for Li$_2$CuO$_2$
parameters (see Fig.~\ref{wave1}.a).
First, we note that the d-intersite ZRS peak has moved
higher in energy from the $d^9$ and core-hole ground states,
lying at a photon energy transfer $\Omega$ of 3.1 eV, and
resonant at a higher incident photon energy $\omega_i -
\epsilon_{4p-1s}=3.1$ eV. The bonding- anti-bonding splitting is
smaller and the resonant peaks are more sharp. In addition, the
CT $(d^{10}L)$ excitations move higher in energy, lying at
$\omega_i - \epsilon_{4p-1s}=3.9$ eV and $\Omega=6.2$ eV. This
energy cost arises from the higher oxygen site energies as well
as the reduction in exchange energy gain with the Cu spin.

\begin{figure}[t]
\includegraphics*[width=\columnwidth,angle=0]{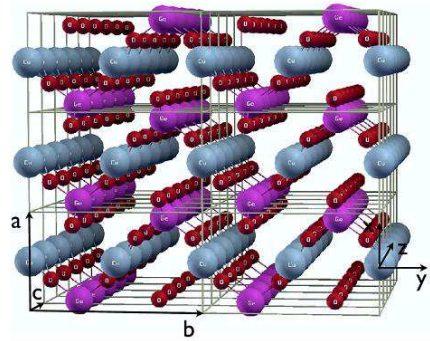}
\caption{CuGeO$_3$ structure.} \label{Structure}
\end{figure}

\begin{figure}[t]
\includegraphics*[width=\columnwidth,angle=0]{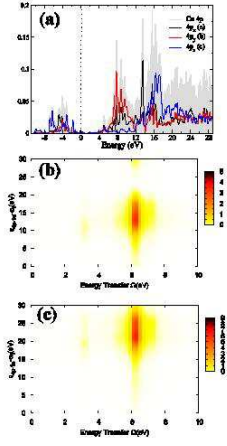}
\caption{{\bf (a)} Cu-$4p$ partial density of states in CuGeO$_3$ calculated
  in WIEN2k.
{\bf (b)} RIXS spectrum for parameters corresponding to
CuGeO$_{3}$, convoluted with the $p_{y}$-DOS.
{\bf (c)} RIXS spectrum for parameters corresponding to
CuGeO$_{3}$, convoluted with the $p_{z}$-DOS.} \label{Geconv}
\end{figure}

The CuGeO$_3$ structure in Fig.\ref{Structure} is
different from the one of Li$_2$CuO$_2$:\ the copper chains
run now along the c-direction and they are tilted from the main-axis
planes.
We present in Fig.~\ref{Geconv}(a) the $4p$ projected DOS for CuGeO$_3$.
The coordinates system for which the calculation has been done
corresponds to the one presented in Fig.\ref{Structure}.
As one can see, the result is strongly dependent on polarization:
the density of $4p_{y}$ is peaked
around $\sim 8$ eV, the $4p_{z}$-density has its stronger weight
around $\sim 18$ eV and is much broader.

The corresponding RIXS results are shown for photon polarizations
along $p_{y}$ and $p_z$ in Fig.~\ref{Geconv}(b) and (c), respectively.
The overall shape of the spectra
does not change much compared to the unconvoluted results (see Fig.~\ref{Ge})~:
one can still identify clearly a low-energy Zhang-Rice peak and a brighter
$d^{10}L$ peak at higher energy as well as the poorly screened $d^9$
excitation peak. The main difference is a shift on the incident photon
energy axis, of $\sim 12$ eV for $p_{y}$ and $\sim 21$ eV for
$p_{z}$.

\begin{figure}[t]
\includegraphics*[width=\columnwidth,angle=0]{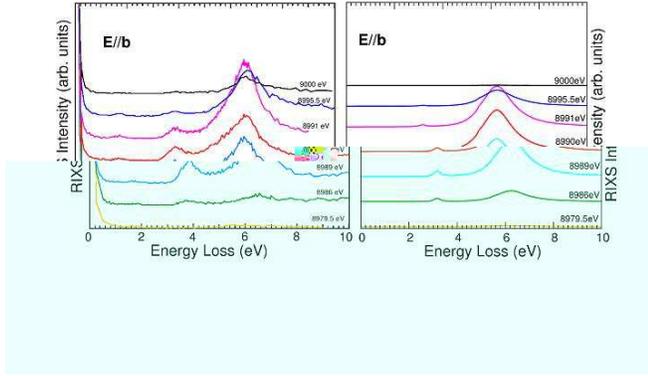}
\caption{{\it (Left)} Experimental RIXS spectrum for CuGeO$_3$.
{\it (Right)} Theoretical spectrum obtained for the same
polarization.} \label{water_Ge}
\end{figure}

\begin{figure}[t]
\includegraphics*[width=\columnwidth,angle=0]{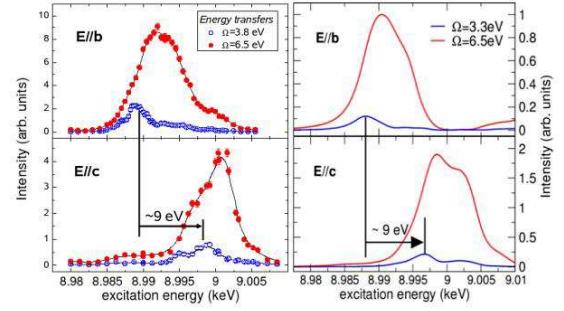}
\caption{{\it (Left)} Experimental RIXS spectrum for CuGeO$_3$ with
polarization dependence.
{\it (Right)} Exact-diagonalization results obtained after
convolution with the $4p$ projected DOS of Fig.\ref{Geconv}.(a)
(see Subsection \ref{convo})
} \label{Ge_exp}
\end{figure}

These results are compared to the recent experimental spectrum
obtained by Hill {\it et al.}\cite{Geck} in Fig.~\ref{water_Ge}
and \ref{Ge_exp}. In this experiment a CT $(d^{10}L)$ peak at
$\Omega\approx 6.5$ eV and the ZRS peak as well, closer to the
elastic line, at about $\sim 3.8$ eV, were identified. These two
observed peaks correspond to the two peaks seen in Fig.~\ref{Ge}.

In Fig.~\ref{water_Ge} we have used an energy-dependent damping
for the Lorentzian representing the delta function of
Eq.(\ref{rixs-formula}). This energy dependence of the line width
accounts for the fact that the higher energy states lie in a high
density continua. For $\hbar\Omega < 5.5$eV, a quadratic energy
dependence of the damping rate was taken and assumed to saturate
at higher Raman shifts. This form was chosen in order to account
for the relative intensity and width of the ZR and $d^{10}L$
peaks.

We note that the slight discrepancy between theory and experiment
in the position and intensity of the d-intersite ZRS may be
remedied by fine-tuning the cluster parameters, to more
accurately reflect the material properties (such as $J$ and the
effective ZR-hopping $t$), as well as incorporating a more
realistic resolution broadening parameter. Another difference
between theory and experiment is the relative intensity of the
peaks for different polarizations: while the peaks are more
intense for a polarization along ${\bf b}$ in the experiment, it
is the opposite for the simulation. This issue may be related to
the two following points: (i) from a theoretical point of view the
relative intensity of the peaks is sensitively linked to the
relative peaks height of the DOS (since in Eq.(1) it appears to
be squared), thus a slight variation in the DFT calculation might
make a difference. (ii) Due to experimental uncertainties a
direct comparison of the measured absolute intensities for ${\bf
e}\|{\bf b}$ and ${\bf e}\|{\bf c}$ is not possible. However, for
a fixed polarization the measured relative intensities of the ZR
and the CT feature can be compared.


Finally we remark that the crystal structures of CuGeO$_3$ and
Li$_2$CuO$_2$ are strongly different (see for instance
Refs.~\onlinecite{Braden,Sapina}). In both cases the CuO$_2$
chains run along a principal axis of the crystal but in CuGeO$_3$
they are not oriented along the same plane, whereas for
Li$_2$CuO$_2$ they are. This has strong implications concerning
the interpretation of the experimental results~: for CuGeO$_3$ the
polarization-dependent experiments will give RIXS spectra which
are a result of convolution with the projected DOS in a rotated
basis. The result of such a convolution is given in the right
panel of Fig.\ref{Ge_exp}.

\section{Conclusions}
A full exact-diagonalization treatment of multi-band Hubbard clusters has been
used to calculate the polarization, incident photon energy,
and transfered photon energy dependencies, of RIXS in one-dimensional,
edge-shared insulating cuprates. By building upon previous work,
\cite{vanveenendaal} the RIXS spectrum can be constructed using
XPS final-states as the intermediate states of inelastic x-ray
scattering. The intensity and resonance profile, as
well as the identity and character of the main RIXS peaks has been
obtained using the eigenstates and eigenenergies of clusters with
and without the core-hole. These peaks correspond to
non-locally screened d-intersite ZRS $d^9L$, local CT $d^{10}L$, and
poorly screened $d^9$, each having intensities and resonance profile
determined by the overlap of these states with the intermediate core-hole
states.

The polarization dependent RIXS spectra were obtained by
convolving the raw RIXS spectra with the copper $4p$ projected
DOS determined by WIEN2k density functional calculations. While
the resulting RIXS spectra includes the $4p$ states as merely
spectators to the re-arrangement of copper-oxygen valence charge
density, a polarization and material dependent RIXS emerges due
to the peak structures and intensities of the $4p$
orbitally-projected intermediate states.

Direct comparison (see Figs.~\ref{Li_comp},\ref{Ge_exp}) with recent data for
two edge-shared cuprates,
Li$_2$CuO$_2$ and CuGeO$_3$,\cite{LiCuO,CuGeO,CuGeO_recent,Suga,both}
shows both qualitative and, more importantly, quantitative agreements
between theory and experiment
for the shape, energy, width, and resonance profile of RIXS spectral
features. This indicates that the local
cluster model can adequately reproduce the data for photon energy
losses in the energy range of 2-7 eV. We again remark that $d-d$
excitations are not captured in the model, and further, that
spin-flip excitations, which require large clusters to observe
bi-magnon spin rearrangements at long wavelengths, are also very
weak. Better agreement of the calculated spectra with experiment may
be obtained by using more refined cluster parameters, multi-pole
couplings, and more accurate representation of resolution broadening.

A polarization dependent RIXS experimental study, for Li$_2$CuO$_2$,
should find significant differences in the resonant profiles for
in-plane polarization and polarization perpendicular to the
CuO$_2$-plane. Fig.~\ref{Liconv} indicates that peak intensities are
brighter and their resonance energies are shifted by $\sim 10$ eV for in-plane
compared to perpendicular polarizations. These predictions as well as
others, like multiple resonant features resulting from $4p$-projected
states, open a way to infer from Cu $K-$edge and other indirect RIXS
experiments the unoccupied DOS. In this way RIXS can be seen as a
complementary tool to core-level spectroscopies such as XPS and XAS,
providing at the same time diverse and important
spectral information on elementary
excitations in strongly correlated materials.

In summary, we have explored the Cu $K-$edge spectrum for
different edge-shared copper oxide systems, and have shown that
the energy scales, peak intensities, and resonance profiles can
be explained qualitatively within the context of exact
diagonalization studies of the multi-band Hubbard model. The
prospect of mapping out the full momentum,
and resonance profile dependences would allow for a direct
spectral check of the many roles played by local electronic
correlations in the multi-band Hubbard approach. This remains a
topic of future research.

\acknowledgements

The authors wish to thank J. P. Hill, Y.-J. Kim, Z.-X. Shen, Z.
Hussain, M. Z. Hasan, M. Greven, J. Hancock, and M. Gingras for
many useful discussions. TPD and GAS would like to acknowledge
support of this work in part by NSERC, CFI, ONR Grant
N00014-05-1-0127 (TPD), PREA (TPD), the Alexander von Humboldt
Foundation (TPD) and CIFAR (GAS). JG gratefully acknowledges the
financial support by the DFG. TPD wishes to thank the Pacific
Institute for Theoretical Physics for their hospitality. FV would
like to thank financial support from CIFAR.


\begin{thebibliography}{99}
\bibitem{nb}
Due to the one-dimension geometry of the clusters, especially for
the corner-shared case, the $d^9L$ is not truely a ZRS as it has
more weight on the central, bridging oxygen. However, the
symmetry of this state remains ZR-like, maintaining the ZR signs
of the ligands.
\end{thebibliography}
\end{document}